\documentclass[conference] {IEEEtran}
\IEEEoverridecommandlockouts
\usepackage{cite}
\usepackage{amsmath,amssymb,amsfonts}
\usepackage{algorithmic}
\usepackage{graphicx}
\usepackage{textcomp}
\usepackage{xcolor}
\usepackage[fontsize=11pt]{fontsize}
\usepackage[hidelinks]{hyperref}
\usepackage{siunitx}
\usepackage{float}
\usepackage{enumerate}
\usepackage{tikz}

\def\BibTeX{{\rm B\kern-.05em{\sc i\kern-.025em b}\kern-.08em
    T\kern-.1667em\lower.7ex\hbox{E}\kern-.125emX}}
    
\newcommand\copyrighttextFirst{%
  \footnotesize \textcopyright 2023 IEEE. Personal use of this material is permitted. Permission from IEEE must be obtained for all other uses, in any current or future media, including reprinting/republishing this material for advertising or promotional purposes, creating new collective works, for resale or redistribution to servers or lists, or reuse of any copyrighted component of this work in other works.}
\newcommand\copyrightnoticeFirst{%
\begin{tikzpicture}[remember picture,overlay]
\node[anchor=north,yshift=-10pt] at (current page.north) {\fbox{\parbox{\dimexpr\textwidth-\fboxsep-\fboxrule\relax}{\copyrighttextFirst}}};
\end{tikzpicture}%
}

\newcommand\copyrighttextSecond{%
  \footnotesize \textcopyright This paper has been accepted for publication in the 25th Conference on Power Electronics and Applications (EPE'23). Please cite the paper as: J. M. Sanz-Alcaine, F. J. Perez-Cebolla, C. Bernal-Ruiz, A. Arruti and I. Aizpuru, J. Sanchez, "Loss Measurement of Low RDS Devices Through Thermal
Modelling - The Advantage of Not Turning it Fully On", 25th Conference on Power Electronics and Applications (EPE'23), 2023.}
\newcommand\copyrightnoticeSecond{%
\begin{tikzpicture}[remember picture,overlay]
\node[anchor=south,yshift=10pt] at (current page.south) {\fbox{\parbox{\dimexpr\textwidth-\fboxsep-\fboxrule\relax}{\copyrighttextSecond}}};
\end{tikzpicture}%
}

\begin{document}

\title{
\LARGE
Loss Measurement of Low $R_{DS}$ Devices Through Thermal Modelling - The Advantage of Not Turning it Fully On

\vspace{0.5em}
\begin{center}
{\large $\text{José Miguel Sanz-Alcaine}^1$, $\text{Francisco Jose Perez-Cebolla}^1$, $\text{Carlos Bernal-Ruiz}^1$, $\text{Asier Arruti}^2$, $\text{Iosu Aizpuru}^2$, $\text{Juan Sanchez}^3$

$^1$ Group of Power Electronics and Microelectronics (GEPM), Aragón Institute of Engineering Research (I3A), University of Zaragoza, Zaragoza, Spain\\
$^2$ Mondragon Unibertsitatea, Galarreta Campus, Hernani, Spain\\
$^3$ Infineon Technologies Austria AG, Villach, Austria\\
E-Mail: jm.sanz@unizar.es, fperez@unizar.es  
}
\end{center}

\vspace{-2em}
}
\maketitle
\copyrightnoticeFirst

\copyrightnoticeSecond
\vspace{-3em}

\section*{Acknowledgment}
This work has been supported by the Spanish project CDTI - MIG-20201042 and a DGA PhD grant. In addition, this project is supported by the KDT Joint Undertaking under grant agreement no 101096884 and Austria, Belgium, Czech Republic, Denmark, Germany, Greece, Netherlands, Norway, Slovakia, Spain, Sweden and Switzerland.
Funded by the European Union. Views and opinions expressed are however those of the author(s) only and do not necessarily reflect those of the European Union, KDT JU or the national granting authorities. Neither the European Union nor the granting authorities can be held responsible for them. 

\begin{figure}[h]
    \centering
    \includegraphics[width=0.2\textwidth]{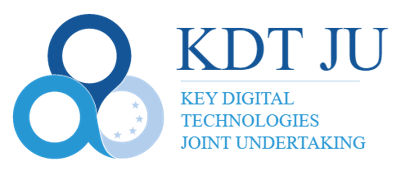}
    \qquad
    \includegraphics[width=0.12\textwidth]{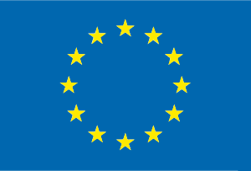}
    \label{fig:enter-label}
\end{figure}

\vspace{0.4cm} 
\begin{IEEEkeywords}
Device characterization, Power losses, Switching and conduction losses, Junction temperature, MOSFET
\end{IEEEkeywords}
\vspace{0.4cm} 

\begin{abstract}
This paper presents and evaluates a novel method for generating power losses on transistors avoiding high currents. These could heat up the circuit tracks, affecting the accurate thermal modeling of the system. The proposed procedure is based on the transistor current regulation with low gate voltages and the linearity between power and temperature, being useful for all transistor technologies (Si, SiC and GaN). Through this method, low DC currents are enough to bring transistors to their thermal limits. Thermal stability issues and their differences between technologies are discussed and an experimental validation of the method is carried out. 
\end{abstract}
\vspace{0.4cm} 

\section{Introduction}
The development of next-generation semiconductors, both wide
bandgap (WBG) \cite{IntroGaN} and new silicon structures \cite{IntroSi}, has opened the window to new paradigms in power electronics in terms of switching speed and heat management. These speeds have made the characterization of devices by electrical methods unfeasible as they are invasive and have insufficient bandwidth \cite{LimBW}. As a result, loss measurement methods based on thermal models have emerged to effectively address this challenge \cite{Kolar2017,Weimer2021}. However, these methods require a previous DC calibration. For low $R_{\text{DS}}$ devices this calibration is performed with high current values that heat up the PCB tracks, which affects the thermal modelling of the system. 
In order to avoid this unwanted effect and to improve the accuracy of the power loss determination, in the proposed procedure, the transistor is forced to operate in the saturation region. Thus, with the channel not fully created, it is possible to generate large losses with very low currents, thereby isolating the transistor from the circuit. The method is valid for both WBG and Si devices. The theoretical foundations of the method are presented in the work, and experimental results are shown to validate them.

The Double-Pulse Test (DPT) \cite{DPTReview} is a widely recognized and accepted technique for determining switching losses by measuring the electrical voltage and current of the device during a turn-on and turn-off transient. However, for fast-switching transients in the nanosecond range, which are commonly achieved with WBG devices, the accuracy of the measured soft-switching losses is unsatisfactory \cite{DPTproblems}. Thermal measurements, based on calorimeters, for characterization of power losses were introduced decades ago for characterization of RF semiconductors and electrical machines \cite{CaliRF,CaliMotores}. This technique has been in continuous development throughout the years, achieving lower characterization time \cite{Guacci2020} and enabling to segregate individual losses \cite{Anurag2020} and \cite {Yamaguchi2021}. For this purpose, a model that relates the temperature of the device to the dissipated power must be obtained. This model can be static or dynamic depending on the final target and, despite the nonlinearities electrothermal effects of transistors, the relationship between power and temperature is linear. Therefore, it can be modeled statically with resistive elements \cite{Kohlhepp2021}, or dynamically, also considering capacitors \cite{Guacci2020}. Likewise, these models can be either applied in discrete devices or in power modules \cite{SiCmodules}. In any case, for both models extraction, a DC current is used by its simplicity of measurement and a least square algorithm is applied for fitting the thermal results \cite{Koch2019}. The smaller the current needed for this calibration, the better accuracy will be reached for the model, as PCB tracks are not excited. For GaN-HEMT devices \cite{GaNreview}, a low current method has already been proposed \cite{Kohlhepp2021}. It is based on the fact that this technology does not include a p-n diode, but can conduct in a way similar to that of a diode in the reverse direction. When current is forced into the source of a ``OFF'' device, a voltage drop is created from the source to the drain and losses occur. However, this method is only applicable to this technology. On the other hand, a prolonged negative voltage applied to the gate could led to damaging the device \cite{GaNdeterioroVth1,GaNdeterioroVth2} and therefore a posterior loss of accuracy in the power estimation. 
However, since the method proposed in this work forces the transistor to operate in the saturation region, these drawbacks are overcome for Si, GaN and SiC. This gives it a universal character. 

\section{Electro-thermal effect on transistors}
In order to fully understand the origin of the losses and the proper characterization procedure, in this section the electro-thermal behavior of the different transistor technologies is reviewed.
\subsection{Silicon}
The instability phenomenon in Si power MOSFETs \cite{StefanoPhD} can be described by taking a careful look at the SPICE level-1 equation for the saturation region

\begin{equation}\label{eq:1}
    I_{\text{D}}(T)=\frac{1}{2}\mu_\text{n}(T)C_{\text{OX}}\frac{W_{\text{cell}}}{L_{\text{ch}}}{[V_{\text{GS}}-V_{\text{th}}(T)]}^2
\end{equation}
where the temperature dependencies of both the electron mobility $\mu_\text{n}$ and the threshold voltage $V_{\text{th}}$ are highlighted. The parameters $C_{\text{OX}}$, $W_{\text{cell}}$ and $L_{\text{ch}}$ represent the gate oxide capacitance, the channel width and the channel length, respectively. For a defined operating condition, i.e. for a fixed $V_{\text{GS}}$, both the channel mobility and the threshold voltage decrease for increasing temperatures determining two counteracting effects on the drain current. Thus: 
\begin{enumerate}[i]
  \item One If the reduction of the mobility dominates over the decrease of the threshold voltage, then the drain current in (\ref{eq:1}) diminishes and the so-called \textit{temperature coefficient of the current} $\alpha$, defined in (\ref{eq:2}), becomes negative.

\begin{equation}\label{eq:2}
  \alpha=\frac{dI_\text{D}}{dT}  
\end{equation}
  \item Two On the contrary, if for increasing temperatures the reduction of the threshold voltage dominates the decrease of the mobility, then $I_{\text{D}}$ increases determining in this case a positive $\alpha$. Those two counteracting mechanisms define two different $V_{\text{GS-ranges}}$ which can be noticed in the temperature dependent transfer characteristics reported in Fig. \ref{fig:TCP}. 

Generally, at reduced current levels, i.e,  for low $V_{\text{GS}}$ voltages, the latter mechanism is dominant. On the other hand, at higher current levels, i.e. at higher $V_{\text{GS}}$ voltages, the former mechanism prevails on the latter one. The boundary between both ranges is determined by the ($V_{\text{GS}}$, $I_{\text{DS}}$) point at which the reduction in mobility with temperature is offset by the reduction in the threshold voltage, resulting in a drain current that is highly insensitive to temperature. For this reason, the pair given by $V_{\text{TCP}}$ and $I_{\text{TCP}}$ in Fig. \ref{fig:TCP} defines the so called Temperature Compensation Point (TCP). Due to the fact that for this point, $\alpha$, is equal to zero, the TCP is often also referred to as the Zero Temperature Coefficient (ZTC) point.

\begin{figure}[h]
    \centering
    \includegraphics{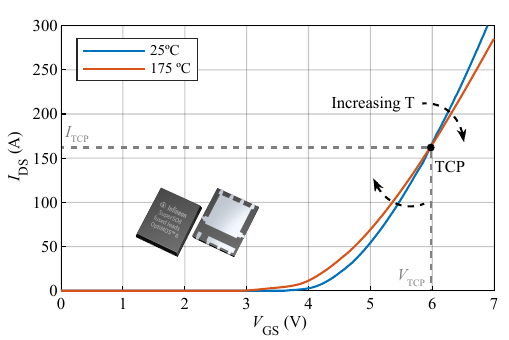}
    \caption{Drain current \textit{vs.} gate voltage at different junction temperatures and definition of the TCP. Two distinct operating regimes can be highlighted: a thermally unstable one for $V_{\text{GS}}<V_{\text{TCP}}$ (or $I_{\text{D}}<I_{\text{TCP}}$) and a thermally stable one for $V_{\text{GS}}>V_{\text{TCP}}$ (or $I_{\text{D}}>I_{\text{TCP}}$).}
    \label{fig:TCP}
\end{figure}

For $V_{\text{GS}}$ lower than $V_{\text{TCP}}$ (and larger than $V_{\text{th}}$), $\alpha$ is positive. This means that a local increase of temperature also determines a local increase of current. Consequently, an increase in current leads to a local increase in power within the cell, which induces a further temperature rise, resulting, in turn, in positive feedback between the current and the temperature increase. If such a mechanism takes place within the device for a sufficient time, extremely high temperatures may be reached within the silicon, and these may lead the device to thermal instability. Therefore, this kind of operation is typically referred as the thermally unstable regime and the related effect is commonly called thermal instability. 
  
\end{enumerate}

\subsection{Silicon Carbide}
\begin{figure}[h]
    \centering
    \includegraphics{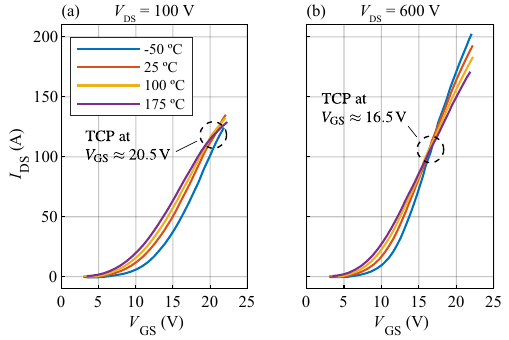}
    \caption{Pulsed transfer characteristics in the conventional temperature range. $T$ = -50 $^{\circ}$C to 175 $^{\circ}$C at (a) $V_{\text{DS}} = \qty{100}{\V}$ and (b) $V_{\text{DS}} = \qty{600}{\V}$. A shift of the TCP is observed \cite{SiCstability}.}
    \label{fig:sic}
\end{figure}

In the case of SiC MOSFETS, the thermal stability is more complex than in Si devices \cite{SiCstability}. In the temperature range from -50 to 175 ° C, in the saturation region, an increase in the temperature forces a significant decrease in $V_{\text{th}}$ and consequently an increase in the current through the device. As a result, the device is thermally unstable over a very wide range of $V_{\text{GS}}$ (Fig. \ref{fig:sic} (a)). For higher temperatures, the thermally stable $V_{\text{GS-range}}$ increases and the TCP moves to lower. This effect becomes especially evident with higher values $V_{\text{DS}}$ (Fig. \ref{fig:sic} (b)). 

\subsection{Gallium Nitride}
\begin{figure}[h]
    \centering
    \includegraphics{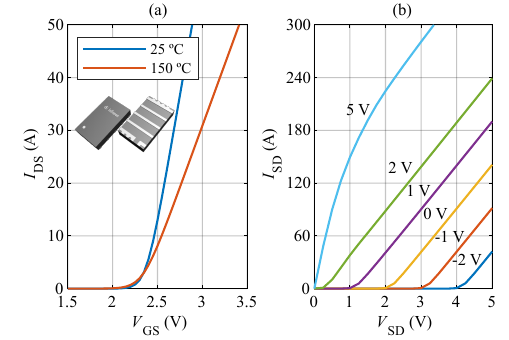}
    \caption{Transfer characteristics (a). Body‐diode forward drop versus source‐drain current (b). Information from \qty{100}{\V}, \qty{3.3}{\mohm} CoolGaN \cite{GaNinf}.}
    \label{fig:ganInf}
\end{figure}

For medium voltage GaN devices \cite{GaNBook}, the temperature changes its behavior in an earlier voltage range than in previous technology, as can be seen in Fig. \ref{fig:ganInf} (a). However, its thermal behavior remains similar to that of silicon. On the other hand, enhancement‐mode GaN transistors do not have a p–n diode, but they operates in a similar a way to a diode in the reverse direction, as discussed before. 

Fig. \ref{fig:ganInf} (b) shows how this ‘body diode’ forward voltage drop varies with the source to drain current and the gate to source voltage. This is the state-of-the-art method for producing calibration power losses in GaN. It is worth noting that the reverse conduction path is created by activating the two-dimensional electron gas (2DEG) in the opposite direction, which involves using a positive gate-drain voltage to enhance the channel. As a result, if the gate voltage is reduced below \qty{0}{\V}, the reverse conduction voltage will rise by the same amount. For example, if the gate drive of a circuit turns off the GaN transistor by applying \qty{0}{\V} to the gate, the $V_{\text{SD}}$ at \qty{0.6}{\A} will be \qty{2}{\V}. If the gate drive of a circuit turns off the GaN transistor by applying a negative \qty{1}{\V} to the gate, the $V_{\text{SD}}$ at \qty{0.6}{\A} will be \qty{3}{\V}. Since the reverse conduction in a GaN transistor is due to turning on the 2DEG, the forward voltage drop will increase with temperature in much the same way as $R_{\text{DS(on)}}$ changes with temperature in an ohmic operation. In contrast, the body diode voltage drop in an Si MOSFET decreases with temperature. 

\section {Proposed solution}
It is evident that all transistor technologies exhibit thermal instability effects in the saturation region. However, this phenomenon only occurs in the absence of a current limitation for $\alpha>0$ or a voltage limitation for $\alpha<0$, while the voltage is fixed by the gate drive circuit. Therefore, by enforcing current regulation and working under $V_{\text{TCP}}$ thermal instability is avoided.

\begin{figure}[h]
    \centering
    \includegraphics[width=\linewidth]{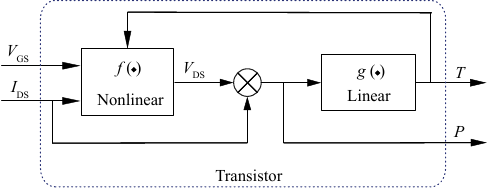}
    \caption{Block diagram of the transistor behaviour. Differentiation and independence from the non-linear system and the linear one is exhibit.}
    \label{fig:BloqueLinealidades}
\end{figure}

\begin{figure*}[hbt!]
\centerline{\includegraphics[width=0.94\textwidth]{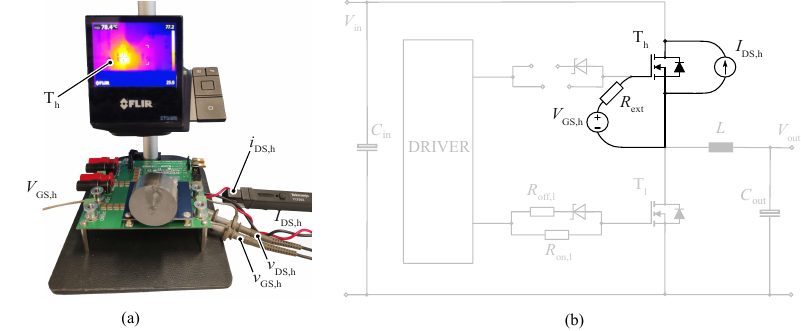}}
\caption{(a) Realization of a Synchronous Buck converter on PCB for the OptiMOS™ 6 devices. (b) Circuit schematic of the circuit where the unused components have been blurred.}
\label{fig:EvalConverter}
\end{figure*}

Thus, in this mode, if the operation of the transistor is ensured in the saturation region instead of in the ohmic region, as is usually done, low currents can still yield high power losses on the device independently of its technology.

Therefore, since the goal of the calibration stage in calorimetric methods is to obtain the linear relationship between power and temperature in the transistor, by stabilizing $V_{\text{GS}}$ and $I_{\text{GS}}$ as inputs of the system, it is possible to obtain a map power \textit{vs} temperature for the latter thermal modeling. 

It is important to note that the sing of $\alpha$ affects only the non-linear electrothermal behavior of the transistor and not the linear relationship between power and temperature (Fig. \ref{fig:BloqueLinealidades}).

\section{Experimental work}
To validate the proposed method, an experimental setup was constructed to generate output characteristic curves at low gate voltage ranges, which are not available in datasheets. Taking into account the fact of variations of $V_{\text{th}}$ for different samples of the same device model \cite{Vthvariable} and to obtain meaningful data for later use in the modeling stage, the device under test (DUT) is already soldered to the converter designed to study losses in future works (Fig. \ref{fig:EvalConverter} (a)).

The DUT is a \qty{100}{\V}@ \qty{3}{\mohm} transistor from Infineon OptiMOS™ 6 \cite{Optimos6}. It is located on the high side of a synchronous buck converter topology where the $V_{\text{GS}}$ and $V_{\text{DS}}$ voltages are available for external monitoring or supplying. Gate resistors were removed to prevent current flow to the driver (Fig. \ref{fig:EvalConverter} (b)). To ensure proper temperature measurement, a high-resolution thermal imaging infrared camera is used to measure $T_{\text{h}}$ and $T_{\text{l}}$. The emissivity factor is controlled and maximized with a high-temperature and high-emissivity tape placed on top of the components.

\begin{figure}[h]
    \centering
    \includegraphics{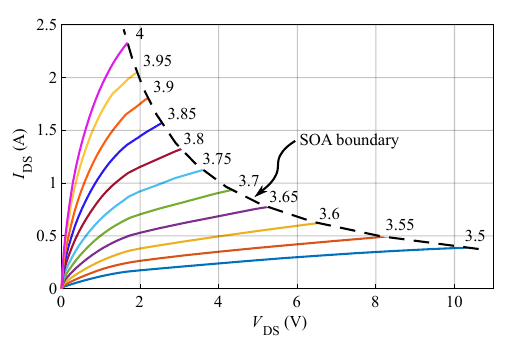}
    \caption{Transistor characteristics obtained experimentally for a set of $V_{\text{GS}}$ values (not included in the datasheet) from 3.5 to \qty{4}{\V} with  increments of \qty{50}{\mV} (solid lines). A power limit curve of \qty{4}{\W} (dashed line) is included. }
    \label{fig:OutputCharacteristics}
\end{figure}

External monitoring and supply of $V_{\text{GS}}$ and $I_{\text{DS}}$ have been carried out by two different programmable power supplies to generate the characteristic curves under low gate voltage conditions. Fig. \ref{fig:OutputCharacteristics} depicts the results, where the dashed line represents the limit of a \qty{4}{\W} safe operating area (SOA) to avoid damaging the transistor.

The information used to create the figure was obtained from voltage and current transients on the DUT. Due to the thermal behavior of operating below the TCP, the curves bend upward over time. Therefore, if current is regulated by the external power supply, an increase in temperature results in a decrease of $V_{\text{DS}}$, leading to a decrease in power and subsequently, a decrement on temperature resulting a thermally stable system. These findings validate the feasibility of generating high power losses through low gate voltages. 

\begin{figure}[h!]
    \centering
    \includegraphics[width=0.48\textwidth]{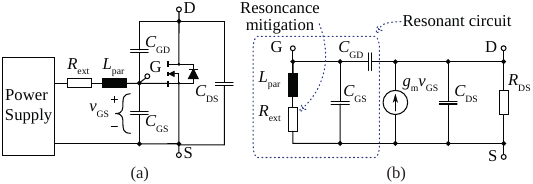}
    \caption{(a) Circuit schematic of the driving section with highlighted parasitic components (b) Circuit schematic of the oscillation effects taking place through the MOSFET small signal model.}
    \label{fig:CircuitoOscilacion}
\end{figure}

The previous data has been obtained in transient operation of the DUT. However, to approach a full characterization of the system, it is of interest to obtain an extent in time losses as the system modeling must thermally reach steady state in all its components. With this target, it has been observed that when driving the DUT from an external supply, a resonance can be produced by the MOSFET parasitics and the cable inductances leading to unstable power losses Fig. \ref{fig:CircuitoOscilacion}. This effect can be mitigated by adding an external resistor that shifts the poles of the oscillating circuit from the imaginary axis.  
 
\begin{figure}[h!]
    \centering
    \includegraphics{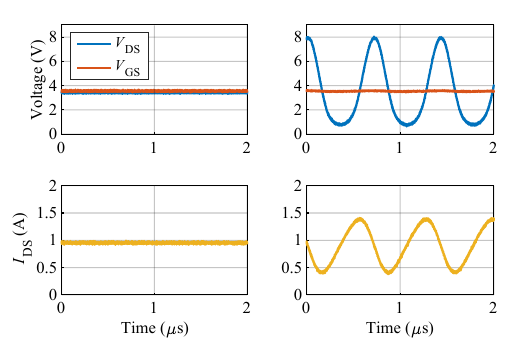}
    \caption{On the left, voltage and current when an external resistance is placed for mitigating the resonance. On the right, voltage and current when it is not placed, forming a \qty{1.4}{\MHz} oscillation.}
    \label{fig:Oscillations}
\end{figure}

Once this external resistor is placed, it can be observed in Fig. \ref{fig:Oscillations} how the current and voltage waveform change from a \qty{1.4}{\MHz} resonance to a static, stable, and easy to measure value.

\begin{figure}[h!]
    \centering
    \includegraphics{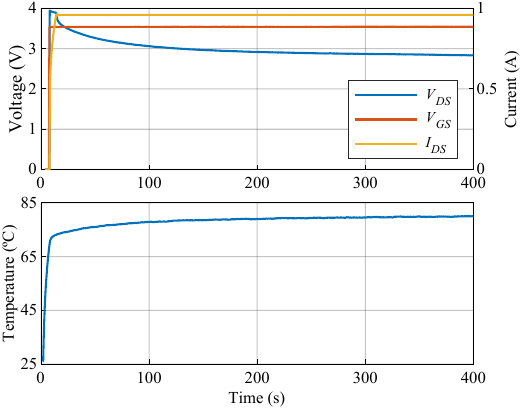}
    \caption{Voltage, current and temperature waveforms measured on the DUT for reaching a thermal steady state.}
    \label{fig:Permanente}
\end{figure}

Once this issue is solved, an example of an operating point is shown as a proof of generating these losses stably. A constant value of $I_{\text{DS}}$ equals to \qty{1}{\A} and  $V_{\text{GS}}$ equals to \qty{3.55}{\V} are used as an example in Fig. \ref{fig:Permanente}. As it can be observed from the data, at the beginning of the process the DUT has not heated up enough for bending the curves of Fig. \ref{fig:OutputCharacteristics} and entering current mode. Once this happens, the voltage $V_{\text{DS}}$ in the transistor, and therefore the power, evolves with the same dynamics from temperature until the desired steady state.  

\section{Conclusion}
The present study proposes a new approach for generating high power losses using low current profiles. The thermal stability of transistors, such as Si, SiC, and GaN, has been briefly explained to support the necessity of operating in a controlled current mode to prevent device overheating. Experimental validations have been conducted, which demonstrate the practicality of the proposed method. The results indicate that, by operating in the saturation region and applying low gate-to-source voltage, stable and high power losses can be achieved. Furthermore, the setup problems have been thoroughly addressed and resolved.

\bibliography{refs} 

\end{document}